# Pressure Effect on the superconducting properties of $LaO_{1-x}F_xFeAs(x=0.11)$ superconductor

W. Lu, J. Yang, X.L. Dong, Z.A. Ren, G.C. Che and Z.X. Zhao

National Laboratory for Superconductivity, Beijing National Laboratory for Condensed Matter Physics, Institute of Physics, Chinese Academy of Sciences, Beijing 100190, China

(*March 28, 2008*)

**Abstract:**

Diamagnetic susceptibility measurements under high hydrostatic pressure (up to 1.03 GPa) were carried out on the newly discovered Fe-based superconductor $LaO_{1-x}F_xFeAs(x=0.11)$. The transition temperature $T_C$, defined as the point at the maximum slope of superconducting transition, was enhanced almost linearly by hydrostatic pressure, yielding a $dT_C/dP$ of about 1.2 K/GPa. Differential diamagnetic susceptibility curves indicate that the underlying superconducting state is complicated. It is suggested that pressure plays an important role on pushing low $T_C$ superconducting phase toward the main (optimal) superconducting phase.

**Introduction**

The newly discovered $LaO_{1-x}F_xFeAs$ superconductor with critical temperature $T_C$ of 26 K has stimulated extensive interest on the layered rare-earth metal oxypnictide family which is free of copper, aiming at pursuing higher $T_C$ or uncovering new evidence on its mechanism [1-8]. The fact that $T_C$ varies, via lattice site substitution, from 4K to 26 K especially the very recently reported 43K [8] further implies a possible rich superconducting phase diagram of this new layered superconductor family of ZrCuSiAs type structure. Besides element substitution, high pressure is often served as an effective parameter to raise $T_C$ by varying carrier concentration through compressing lattice. In addition, high pressure experiments, by changing the interaction between particles, often play very important role on providing new information on the paring interaction and testing proposed theory [9]. Here we report the pressure effect on the newly discovered $LaO_{1-x}F_xFeAs$ superconductor through magnetic susceptibility measurements. The transition temperature $T_C$ was defined as the point where maximum slope of superconducting transition achieves, i.e., the peak of differential diamagnetic susceptibility. $T_C$ was found to increase almost linearly upon hydrostatic pressure increasing, showing a $dT_C/dP$ value of about 1.2 K/GPa. Differential diamagnetic susceptibility curves indicate that the underlying superconducting state is complicated, at least two superconducting transitions coexist. The lower $T_C$ is dramatically enhanced by pressure while the higher one shows little pressure dependence. The role of pressure is hence discussed based on its different effect on the existing superconducting transitions.

**Experimental details**

The F-doped $LaO_{1-x}F_xFeAs$ polycrystalline sample was synthesized by a two-step method. $La_2O_3$, Fe, As, $LaF_3$ powders and La pieces (the purities of all chemicals are higher than 99.99%) were used as starting materials and weighed according to the chemical stoichiometry of $LaO_{1-x}F_xFeAs$. At the first step, the precursor LaFeAs alloy was prepared by argon arc-melting method. The weighed Fe and As powders (here 20% As was added to compensate the volatilization during alloying) were pressed into pellets and put on the water-cooled copper stage together with the La pieces. The mixtures were then melted and alloyed for two minutes. The alloyed ingot became brittle and was ground to fine powder. The second step was performed through a solid-state-reaction process. The prepared LaFeAs powder was mixed with the weighed $LaF_3$ and $La_2O_3$ powders, then ground thoroughly and pressed into pellet. After enclosed in evacuated quartz tubes (with the vacuum better than $10^{-3}$ Pa), the sealed sample was sintered at $1240^oC$ for 50 hours in muffle furnace, with the heating and cooling rates of $100^oC$ per hour. The sample for measurements was processed into approximately 1.34 x 1.1 x 0.4 $mm^3$ rectangular shape.

The structure of the synthesized samples was characterized by high power X-Ray diffraction (XRD, an 18KW MXP18A-HF diffractometer with Cu-$K_\alpha$ radiation). The scanning electronic microscopy (SEM) analysis was performed on a Philips XL30 S-FEG. microscope. The resistivity data were measured by the standard four-probe technique.

The hydrostatic pressure was generated by a commercial pressure cell (Mcell 10) which was especially designed for magnetic property measurement using Quantum Design Magnetic Property Measurement System (MPMS). The pressure was measured in situ by recording the $T_C$ shift of a small piece of Sn included in the Mcell 10. The diamagnetic susceptibilities ($\chi$) were obtained by magnetization measurements carried out during warming cycle under fixed magnetic field after zero field cooling (ZFC) or field cooling (FC) using a Quantum Design MPMS XL-1. Background signals were subtracted simultaneously during the measurements through a commercial functional program. All the data reported here were corrected for the demagnetization factor [10].

**Results and discussion:**

The sample's powder XRD pattern shown in Fig.1a indicated a well indexed tetragonal structure (space group *P4/mmn*) with a = 4.030(1) Å and b = 8.706(2) Å except for weak impurities peaks of LaOF and $FeAs_2$. A SEM image of our sample is presented in Fig. 1b, showing the layered structure feature with the grain size of about 10 micrometers. Fig.2 a is the temperature-dependence of resistivity. A sharp transition starts at 26.3K and zero resistivity is reached when temperature is lowered to 22.3K. The temperature-dependent diamagnetic susceptibility χ measured under 1 Oe after zero-field-cooling is shown in Fig.2 b. The onset $T_C$ of 23K is consistent with the resistivity results, while the 100% magnetic shielding signal evidences bulk superconductivity. All these results indicate the good quality of our sample.

Differential diamagnetic susceptibility was focused to monitor any tiny change in order to investigate the pressure effect on the superconducting properties of this newly discovered superconductor. Shown in Fig. 3 is the differential diamagnetic susceptibility dχ/dT measured at 1 Oe under hydrostatic pressure. $T_C$, defined as the peak of dχ/dT, showed obvious pressure dependence. With pressure increasing from ambient up to 1.03 GPa, Tc raised almost linearly from 20.3K to 21.5 K, yielding a $dT_C/dP$ value of about 1.2 K/GPa (the inset of Fig. 3). To our surprise, the onset $T_C$ showed no obvious change when pressure increased up to 1.03 GPa. In contrast, the superconducting transition, illustrated by the dχ/dT peak in Fig. 3, was narrowed with increasing pressure. In other words, it seems that pressure (below 1.03 GPa) will first improve the superconducting correlation network before interfering the main superconducting phase characterized by onset $T_C$ of 23 K.

In order to clarify the pressure effect on both $T_C$ and superconducting transition width, the pressure dependence of dχ/dT measured at 10 Oe was also investigated, as plotted in Fig. 4. Although the magnetic field of 10 Oe is rather low, the dχ/dT curves were largely broadened and extra features occurred, implying a complicated superconducting state. Two peaks are evident, the temperature of higher peak $T_{PH}$ is around 19.2K while the lower one $T_{PL}$ locates at 16.1K under ambient pressure, indicating the existence of inhomogeneous superconducting state. When pressure increases, the higher peak retains around 19.2K up to 1.03 GPa, showing no obvious pressure dependence. This is consistent with the pressure-insensitive onset $T_C$ mentioned above. However, dramatic pressure dependence occurs for the lower peak. The $T_{PL}$ shift changes from 16.1K under ambient pressure to about 17 K at 1.03 GPa, as shown in the inset of Fig.4, generates a nearly linear pressure dependence with a large slope, i.e., $dT_{PL}/dP$, of about 4 K/GPa (inset of Fig.4). In other words, low $T_C$ phase is more sensitive to pressure, compared with the high $T_C$ phase, which is also the main (optimal) superconducting phase with onset $T_C$ of 23K as illustrated in Fig.2.

Clearly, the onset $T_C$ of optimal superconducting phase showed little pressure dependence,

implying that a pressure up to 1.03 GPa made little effect, by either varying carrier concentration or adjusting inner-and intra-plane interaction via lattice compressing, on the structural frame that bears the optimal superconducting phase. However, notice the fact that the pressure-induced improvement of superconducting transition is obvious as presented above, we still expect the potential of higher $T_C$ triggered by pressure above 1.03 GPa. In addition, the fact that superconducting transition is very sensitive to applied field implies that pressure effect on such as lattice defect (especially in $La_2(O,F)_2$ layer) and grain boundaries as well as other possible weak links might also be considerable.

**Conclusion:**

In summary, we have studied the pressure dependence of diamagnetic susceptibilities of the newly discovered Fe-based superconductor $LaO_{1-x}F_xFeAs$(x=0.11). The transition temperature $T_C$, manifested by the peak of $d\chi/dT$, increased linearly upon hydrostatic pressure, yielding a dTc/dP value of about 1.2 K/GPa. Differential diamagnetic susceptibility curves indicated that the underlying superconducting state is complicated. Pressure was found to play a key role on pushing low $T_C$ superconducting phase toward the main (optimal) superconducting phase. Higher $T_C$ is expected via applying pressure larger than 1.03 GPa.

**Acknowledgement** We thank Ms. S.L. Jia, Y. Gu for their technical assistances. This work is supported by Natural Science Foundation of China (NSFC) and 973 program of China (No. 2007CB925002). We also acknowledge the support from EC under the project COMEPHS TTC.

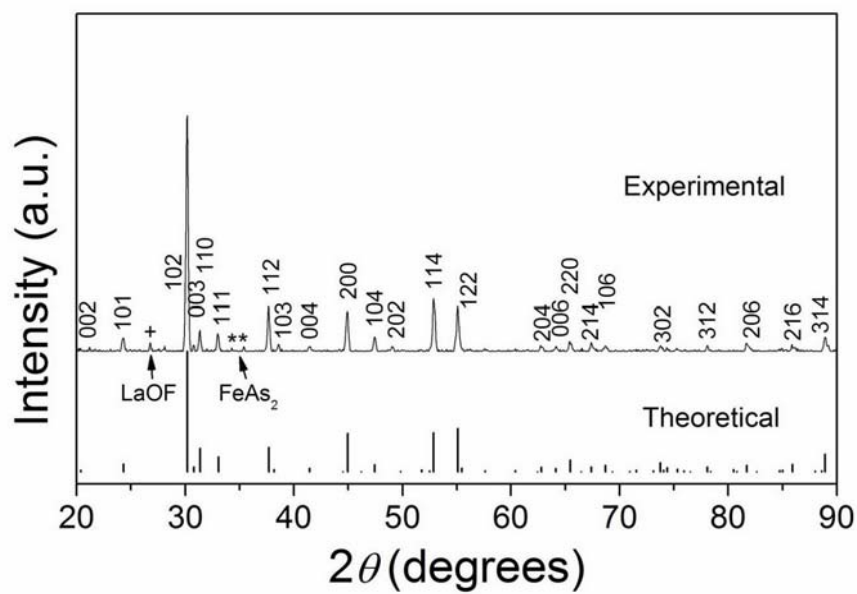

**Fig.1a** Powder XRD patterns of F-doped $LaO_{0.89}F_{0.11}FeAs$. The results from our experiment and theoretical simulation are in good agreement except two additional phases from the impurities LaOF and $FeAs_2$.

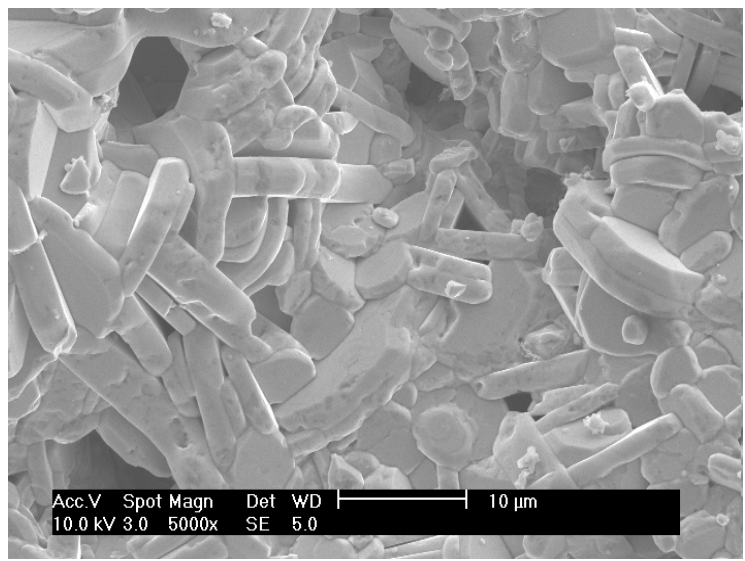

**Fig.1b** SEM image of F-doped $LaO_{0.89}F_{0.11}FeAs$, showing the layered structure feature with the grain size of about 10 micrometers.

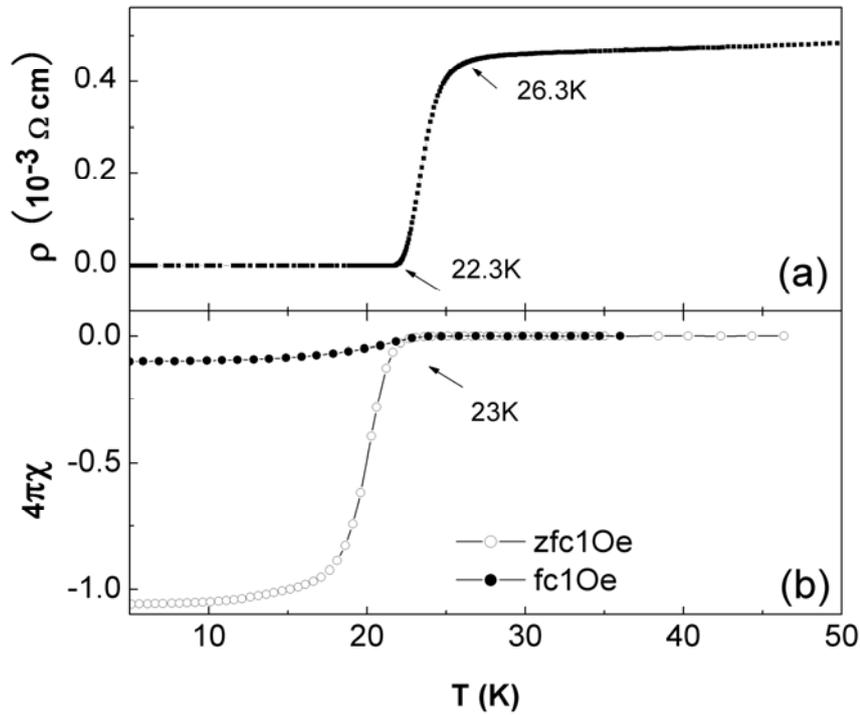

**Fig.2** (a) Electrical resistivity ($\rho$) vs. Temperature. The resistivity begins to drop dramatically below 26.3K and the zero resistivity appears at 22.3K. (b) Temperature dependence of diamagnetic susceptibility measured under 1Oe at ambient pressure, onset $T_C$ is of 23 K.

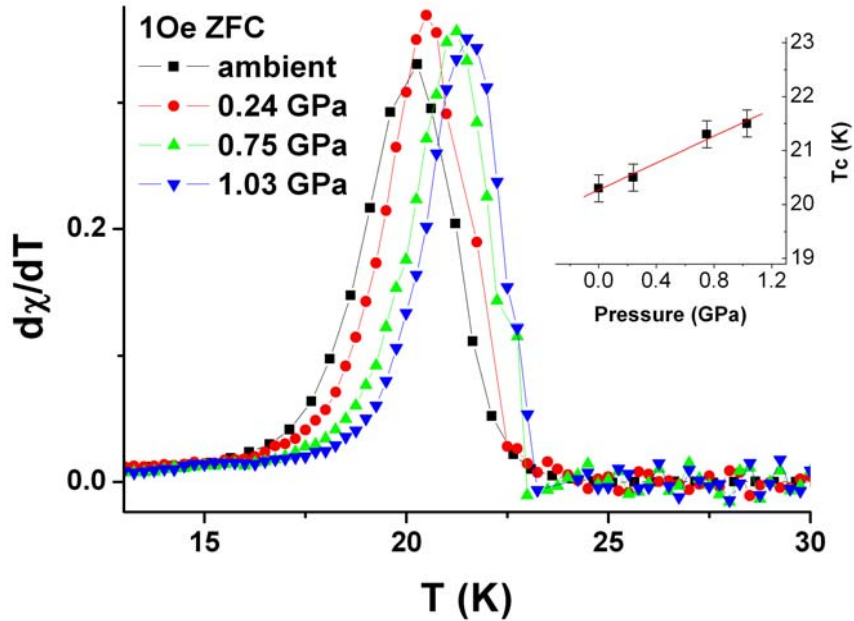

**Fig.3** Differential curves of diamagnetic susceptibility ($d\chi/dT$) measured under 1 Oe after ZFC under different hydrostatic pressures. Inset shows the pressure dependence of $T_C$ defined as the peak of $d\chi/dT$, yielding a $dT_C/dP$ (the slope of red line) of 1.2 K/GPa.

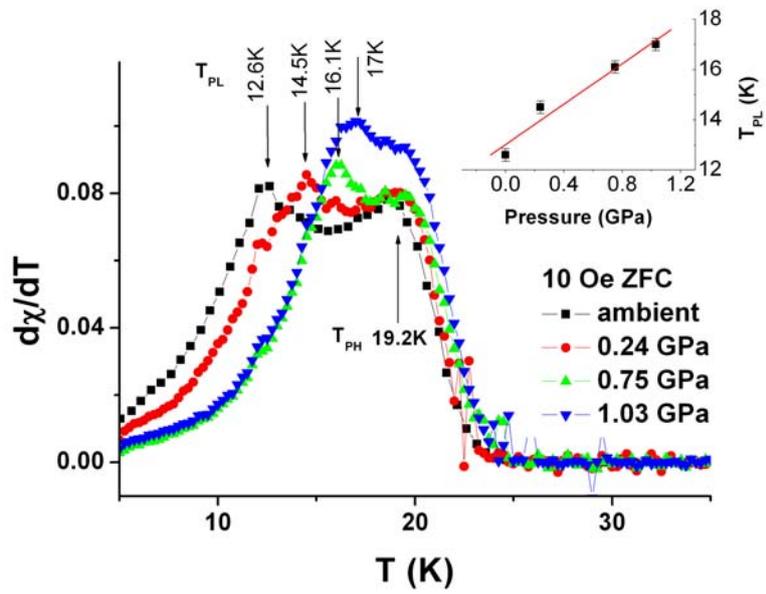

**Fig.4** Differential curves of diamagnetic susceptibility (dχ/dT) measured under 10 Oe after ZFC under different hydrostatic pressures. The low-T peak $T_{PL}$ of dχ/dT shifts toward high temperature with pressure increasing and gives a dTp/dP value of about 4K/GPa (inset).